\documentclass[sigconf]{acmart}

\def\BibTeX{{\rm B\kern-.05em{\sc i\kern-.025em b}\kern-.08emT\kern-.1667em\lower.7ex\hbox{E}\kern-.125emX}}
    
\copyrightyear{2019}
\acmYear{2019}
\setcopyright{acmlicensed}
\acmConference[CIKM '19]{CIKM '19: ACM International Conference on Information and Knowledge Management}{November 03--07, 2019}{Beijing, China}
\acmBooktitle{CIKM '19: ACM International Conference on Information and Knowledge Management, November 03--07, 2019, Beijing, China}
\acmPrice{15.00}
\acmDOI{10.1145/1122445.1122456}
\acmISBN{978-1-4503-9999-9/18/06}
\usepackage{subfigure}

\begin{document}

%
\title{Query-based Interactive Recommendation by Meta-Path and Adapted Attention-GRU}

%

\author{Yu Zhu}
\affiliation{%
  \institution{Alibaba Group}
  \streetaddress{No. 969 Wenyi Road}
  \city{Hangzhou}
  \country{China}}
\authornote{Both authors contributed equally to this research.}
\email{zy143829@alibaba-inc.com}

\author{Yu Gong}
\affiliation{%
  \institution{Alibaba Group}
  \streetaddress{No. 969 Wenyi Road}
  \city{Hangzhou}
  \country{China}}
\authornotemark[1]
\email{gongyu.gy@alibaba-inc.com}

\author{Qingwen Liu}
\affiliation{%
  \institution{Alibaba Group}
  \streetaddress{No. 969 Wenyi Road}
  \city{Hangzhou}
  \country{China}}
\email{xiangsheng.lqw@alibaba-inc.com}

\author{Yingcai Ma}
\affiliation{%
  \institution{Alibaba Group}
  \streetaddress{No. 969 Wenyi Road}
  \city{Hangzhou}
  \country{China}}
\email{yingcai.myc@alibaba-inc.com}

\author{Wenwu Ou}
\affiliation{%
  \institution{Alibaba Group}
  \streetaddress{No. 969 Wenyi Road}
  \city{Hangzhou}
  \country{China}}
\email{santong.oww@taobao.com}

\author{Junxiong Zhu}
\affiliation{%
  \institution{Alibaba Group}
  \streetaddress{No. 969 Wenyi Road}
  \city{Hangzhou}
  \country{China}}
\email{xike.zjx@taobao.com}

\author{Beidou Wang}
\affiliation{%
  \institution{State Key Laboratory of CAD\&CG, Zhejiang University}
  \streetaddress{No.866 Yuhang Road}
  \city{Hangzhou}
  \country{China}}
\email{beidou.wang@gmail.com}

\author{Ziyu Guan}
\affiliation{%
  \institution{Xidian University}
  \streetaddress{No.266 Xingrong Road}
  \city{Xi'an}
  \country{China}}
\email{zyguan@xidian.edu.cn}

\author{Deng Cai}
\affiliation{%
  \institution{State Key Laboratory of CAD\&CG, Zhejiang University}
  \streetaddress{No.866 Yuhang Road}
  \city{Hangzhou}
  \country{China}}
\email{dcai@zju.edu.cn}

%
\renewcommand{\shortauthors}{Zhu et al.}

%
\begin{abstract}
  Recently, interactive recommender systems are becoming increasingly popular. The insight is that, with the interaction between users and the system, (1) users can actively intervene the recommendation results rather than passively receive them, and (2) the system learns more about users so as to provide better recommendation. 
  
  We focus on the single-round interaction, i.e. the system asks the user a question (Step 1), and exploits his feedback to generate better recommendation (Step 2). A novel query-based interactive recommender system is proposed in this paper, where \textbf{personalized questions are accurately generated from millions of automatically constructed questions} in Step 1, and \textbf{the recommendation is ensured to be closely-related to users' feedback} in Step 2. We achieve this by transforming Step 1 into a query recommendation task and Step 2 into a retrieval task. The former task is our key challenge. We firstly propose a model based on Meta-Path to efficiently retrieve hundreds of query candidates from the large query pool. Then an adapted Attention-GRU model is developed to effectively rank these candidates for recommendation. Offline and online experiments on Taobao, a large-scale e-commerce platform in China, verify the effectiveness of our interactive system. The system has already gone into production in the homepage of Taobao App since Nov. 11, 2018 (see https://v.qq.com/x/page/s0833tkp1uo.html on how it works online). Our code and dataset are public in https://github.com/zyody/QueryQR.
\end{abstract}

%
%


%

%

%
\maketitle
\section{Introduction}
\emph{Interactivity} plays an important role in influencing user experience in daily life. For example, compared to watching TV, most children prefer playing with the smartphone or tablet. One main reason is that, children can only passively receive TV shows from TV (with few interactions when they change channels), while they have lots of interactions with the smartphone or tablet when, for example, playing mobile games.

It is the same case with recommender systems (RS). Intuitively, by introducing the interactivity in RS (i.e. users interact with RS), users can actively intervene the recommendation results rather than passively receive them. In addition, the system will learn more about users so as to provide better recommendation. Both will improve the user experience. However, how to utilize the interactivity to improve the recommendation performance has not been well studied in the past decades. 

There is an increasing number of works \cite{chen:2018stabilizing,christakopoulou:2018q} concentrating on interactive RS recently. Personal assistants learn users' preferences by conversing with users \cite{jugovac:2017interacting}. Users' interests can also be mined by questionnaires \cite{elahi:2013personality}. Reinforcement learning \cite{chen:2018stabilizing} /multi-armed bandit \cite{li:2010contextual}/active learning \cite{rubens:2015active} methods focus on balancing the explore-exploit tradeoff in RS. They learn users' preferences by recommending explored items and acquiring their feedback. However, personal assistants and reinforcement learning/multi-armed bandit/active learning methods usually need multi-round interactions to well learn users, thus users who want quick$\&$accurate recommendations would not be satisfied. The questionnaires are often manually generated and not well personalized. These issues prevent the above methods from being the best fit for interactive recommendation tasks.

\begin{figure}[tb]
\begin{center}
\includegraphics[scale=0.21]{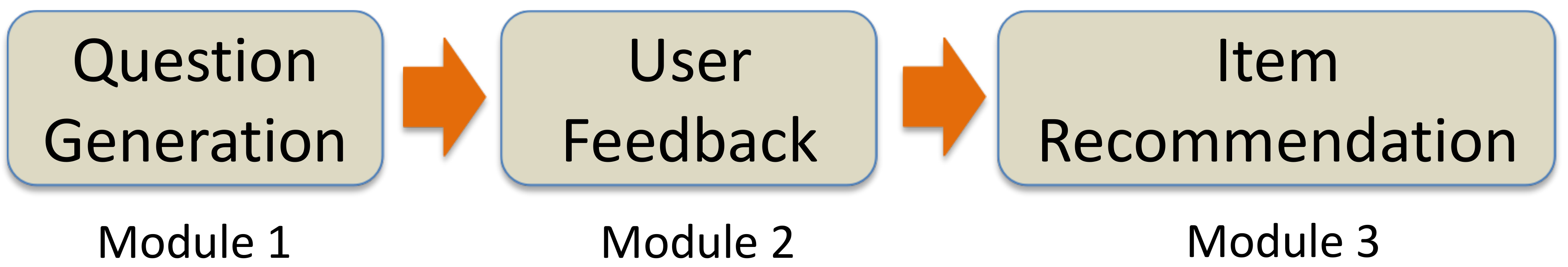}
\end{center}
   \caption{The Question$\&$Recommendation framework.
}
\label{fig:ThreeModules}
\end{figure}
Single-round interaction and automatically generating personalized questions are promising solutions to the aforementioned challenges. A framework for the single-round interactive RS is proposed in \cite{christakopoulou:2018q}, which contains three main modules as shown in Figure \ref{fig:ThreeModules}. Specifically, when a user browses items in RS, the system would generate a question to consult him about what his interest is (Module 1). Once his feedback is obtained (Module 2), the system could then provide more accurate recommendation (Module 3). 

Our work is also based on this framework. Now we describe our design in terms of the three modules, respectively.\vspace{10pt}

\begin{figure}[tb]
\begin{center}
\includegraphics[scale=0.27]{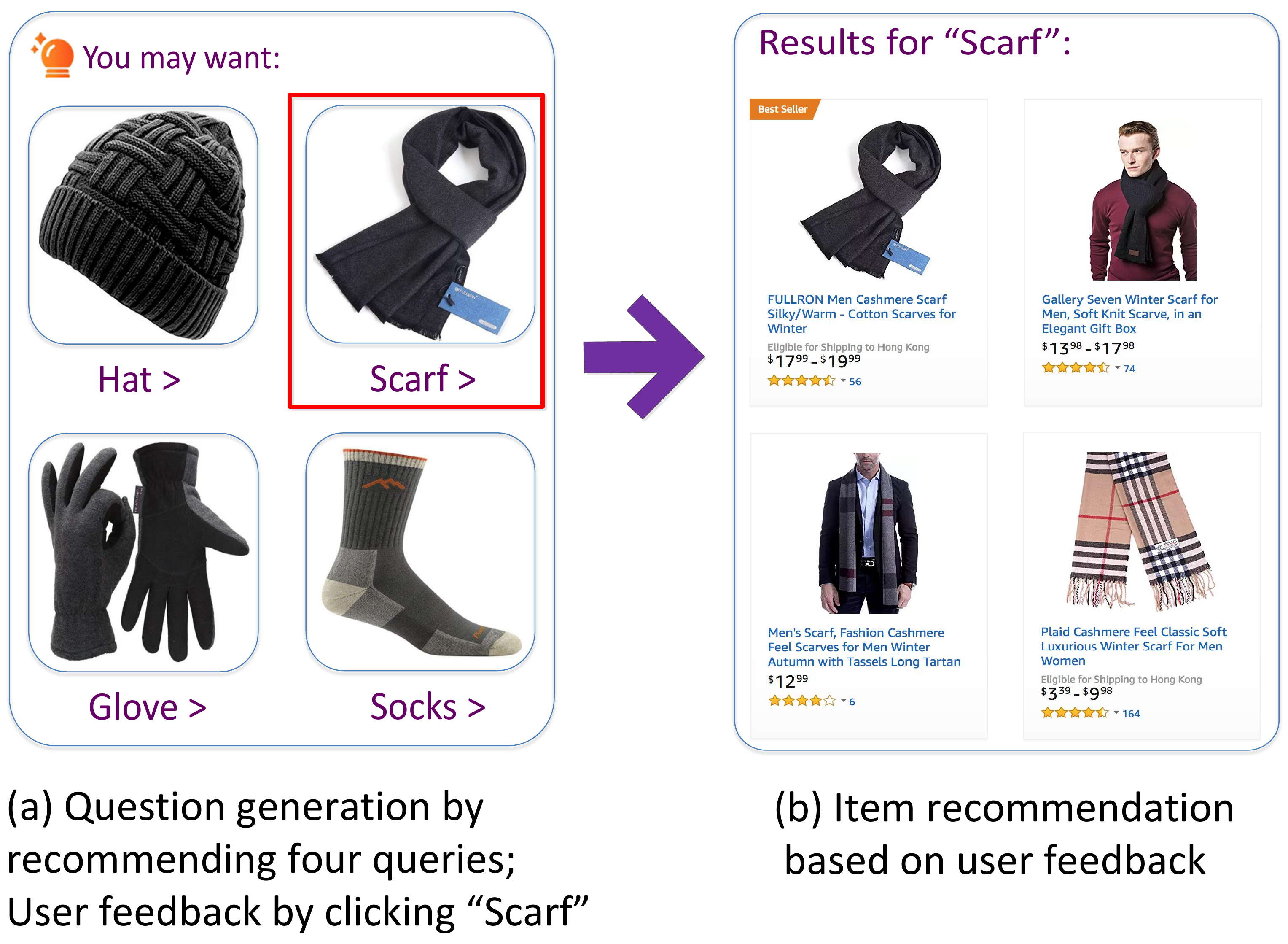}
\end{center}
   \caption{The system generates 4 queries to consult the user about whether he wants a hat/scarf/glove/socks. He answers ``Yes, I want a scarf'' by clicking ``Scarf''. Then personalized item recommendation is performed based on the query ``Scarf''.
}
\label{fig:feedbackRecommend}
\end{figure}

\noindent \textbf{\emph{Question Generation}:} We transform the question generation task into a query recommendation task. For example, in Figure \ref{fig:feedbackRecommend} (a), four queries ``Hat", ``Scarf", ``Glove" and ``Socks'' are recommended to the user, corresponding to the questions ``Do you want to buy a hat/scarf/glove/socks?". We choose to generate questions based on queries due to the following three considerations:


\begin{itemize}
\item Queries are typed by users, in order to find their preferred items. Therefore, queries reflect users' potential preferences.
\item The questions generated in \cite{christakopoulou:2018q} are topic-based. Compared to topics, queries can capture more fine-grained preferences.
\item The search log has plenty of queries. These queries are widely distributed and cover almost all preferences from different users in various circumstances (e.g. different seasons).
\end{itemize}

After filtering queries with low frequency, millions of queries are collected from the search log. To recommend queries (from the large query pool) that can best reflect users' preferences in an efficient fashion, various important information in e-commerce websites should be well exploited, including (1) heterogeneous relations among different objects (users, items, queries, etc.); (2) rich features of these objects, e.g. the text and category (Dresses, Smartphones, etc.) information of items and queries; (3) the sequential information, action types (\emph{click}, \emph{purchase}, etc.) and timestamps of users' behaviors. Inspired by \cite{covington:2016deep}, we split query recommendation into two stages: Candidate Generation and Ranking. Specifically, we firstly propose a model based on Meta-Path \cite{zhao:2017meta} to efficiently generate hundreds of query candidates from the large query pool, with the help of heterogeneous relations. Then an adapted Attention-GRU \cite{chorowski:2015attention} model is developed to effectively rank these candidates, by utilizing all the information described above. In this way, high efficiency and accurate recommendation are both achieved.

\noindent \textbf{\emph{User Feedback}:} As shown in Figure \ref{fig:feedbackRecommend} (a), the user answers ``Yes, I want a scarf" by clicking ``Scarf" or ``No, I do not want them" by ignoring them. 

\noindent \textbf{\emph{Item Recommendation}:} Finally, as shown in Figure \ref{fig:feedbackRecommend} (b), items are recommended based on the clicked query and the user's historical behavior. Actually, this can be seen as a classical personalized retrieval task, and many successful algorithms in the information retrieval area could be adopted. In this way, the recommendation (i.e. the retrieved results) is ensured to be closely-related to users' feedback (i.e. query).

Apart from the motivated example in Figure \ref{fig:feedbackRecommend}, we show some other possible user cases to give a better understanding of our system:
\begin{itemize}
\item After the user clicking some items about trips to Tokyo, he is recommended with queries ``Hotel in Tokyo'', ``Flight to Tokyo'', ``Travelling bag'' and ``Toiletries''.
\item After buying a Nikon camera, he is shown with ``Memory card'', ``Lenses'', ``Camera battery'' and ``Protection cases for camera''.
\item After favoring a pair of shoes, he is shown with ``Nike shoes'', ``Adidas shoes'', ``Sneakers'' and ``Leather shoes''.
\end{itemize}

Since item retrieval algorithms in most e-commerce platforms are mature, we directly adopt them as the models in \emph{Item Recommendation}. Therefore, query recommendation in \emph{Question Generation} is our key challenge. We will focus on it in the rest of our paper. 

This paper's contributions are outlined as follows.
\begin{itemize}
\item We design a novel query-based interactive RS. Compared to state-of-the-art interactive RS, e.g. \cite{christakopoulou:2018q}, our system can accurately generate personalized questions from millions of automatically constructed questions (since we have millions of queries) and item recommendation is ensured to be closely-related to users' feedback, which result in better user experience in interactive RS.
\item We propose a solution by Meta-Path and adapted Attention-GRU for query recommendation. This solution follows a Candidate Generation and Ranking schema. We introduce Meta-Path into the Candidate Generation stage and customize the calculation of meta path scores, so that query candidates can be efficiently generated considering heterogeneous relations and the procedure is more explainable. We introduce Attention-GRU into the Ranking stage and propose two important modifications on Attention-GRU, which significantly improves its ranking performance.
\item We conduct extensive offline and online experiments on a large-scale e-commerce platform, i.e. Taobao. The experimental results (especially the response from online users) prove the effectiveness of our query-based interactive RS.
\end{itemize}
\section{Related Work}
\subsection{Interactive RS and Query Recommendation}
\cite{he:2016interactive,jugovac:2017interacting} are excellent surveys on interactive RS. How we differ from the most  relevant works are described in Introduction. In addition, compared to \cite{christakopoulou:2018q}, our solution in \emph{Question Generation} is based on queries and is carefully designed to improve its efficiency and effectiveness. Meanwhile, we transform \emph{Item Recommendation} into a retrieval task so that we can address it by adopting existing retrieval algorithms.

Query recommendation in most previous works \cite{huang:2016kb,zhao:2015mobile} is to facilitate the search of web pages, locations, etc. They usually exploit information, e.g. searched queries and clicked links, in search logs. Ours is for item recommendation, and is based on both of search and recommendation logs, including some special information in e-commerce websites. Note that our framework is not limited to queries and other objects (e.g. videos) can also be utilized to generate questions. \emph{Item recommendation} will then be based on behaviors on these objects, instead of only the clicked queries. We will explore it in our future work.
\subsection{Meta-Path}
\emph{Meta-Path} \cite{sun:2011pathsim} describes how two nodes in a graph are connected via different types of paths. Specifically, given a directed graph $\mathbf{G} = (\mathbf{V}, \mathbf{E})$, where $\mathbf{V} = \{V_0, V_1,\cdots\}$ is the node set, and $\mathbf{E} = \{E_0, E_1,\cdots\}$ is the edge set. A meta path $P = V_0 \stackrel{E_0}{\longrightarrow} V_1 \stackrel{E_1}{\longrightarrow} \cdots \stackrel{E_{k-1}}{\longrightarrow} V_k$ in $\mathbf{G}$ defines a complicated relation between $V_0$ and $V_k$. Several works \cite{yu:2014personalized,yu:2013collaborative} exploit Meta-Path to improve the performance of RS. Corresponding to RS, entities such as users and items construct the nodes, and relations such as users consuming items are the edges. Many recommendation algorithms can be represented by meta paths. For example, item-based collaborative filtering (Item-CF) \cite{sarwar:2001item} and Content-based recommendation (CBR) \cite{pazzani:2007content} can be represented by meta paths: $p^{Item-CF}/p^{CBR} = u \stackrel{Consume}{\longrightarrow} i \stackrel{Similar}{\longrightarrow} i^\prime$, indicating that user $u$ consuming item $i$ may also prefer a similar item $i^\prime$ (the similarity is calculated by collaborative behaviors for Item-CF and by item contents/attributes for CBR).
Similarly, user-based collaborative filtering (User-CF) \cite{zhao:2010user} and Social-aware recommendation (SR) \cite{tang:2013social} are represented by: $p^{User-CF}/p^{SR} = u \stackrel{Similar}{\longrightarrow} u^\prime\stackrel{Consume}{\longrightarrow} i$, indicating that $u$ may favor what his similar users (the similarity is calculated by collaborative behaviors for User-CF and by social relations for SR) have consumed.
\cite{zhao:2017meta} proposes a Meta-Graph based recommendation method to capture more complex semantics. Since there is no complex semantics in our task, thus we use Meta-Path for simplicity. Compared to previous Meta-Path models, we combine Meta-Path with Attention-GRU, instead of using Meta-Path individually. Thus heterogeneous relations and sequential behaviors are both considered for recommendation.

\subsection{Attention-GRU}
\emph{Attention-GRU} \cite{chorowski:2015attention} refers to GRU \cite{cho:2014learning} with the attention schema \cite{bahdanau:2014neural}. It typically generates an output sequence $y = (y_1,\cdots, y_T)$ from an input sequence $x = (x_1,\cdots, x_n)$, where $x$ is usually encoded to a sequential representation $h=(h_1,\cdots, h_n)$ by an \emph{encoder}. $y_m$ in $y$ is generated by:
\begin{align}
&\alpha_m = Attend(s_{m-1}, h)\label{eq:Attention-GRU1},\\
&g_m = \sum_{k=1}^n\alpha_{mk}h_k\label{eq:Attention-GRU2},\\
&s_m = Recurrence(y_{m-1},s_{m-1},g_m)\label{eq:Attention-GRU3},\\
&y_m\sim Generate(y_{m-1},s_m,g_m)\label{eq:Attention-GRU4},
\end{align}
where $Attend$ and $Generate$ are functions. $s_m$ is the hidden state. $\alpha_m$ is a vector whose entry $\alpha_{mk}$ indicates the \emph{attention weight} of the $k$-th input. $g_m$ is called a \emph{glimpse} \cite{mnih:2014recurrent}. $Recurrence$ represents the recurrent activation. In Attention-GRU, the recurrent activation is GRU. 

RNN solutions for behavior modeling are becoming increasingly popular \cite{hidasi:2015session,hidasi:2016parallel}. The most related work to ours is \cite{zhu:2018brand}. Our main difference lies in (1) we modify the attention schema in Attention-GRU while \cite{zhu:2018brand} does not; (2) Besides behavior features extracted by Attention-GRU, we also incorporate other valuable features (e.g. features from Meta-Path) for recommendation in a non-trivial fashion.

\section{Query Recommendation}
As described in Introduction, query recommendation is our key challenge. Our solution contains two stages: Candidate Generation and Ranking.
\subsection{Meta-Path for Candidate Generation}
\begin{figure}[tb]
\begin{center}
\includegraphics[scale=0.26]{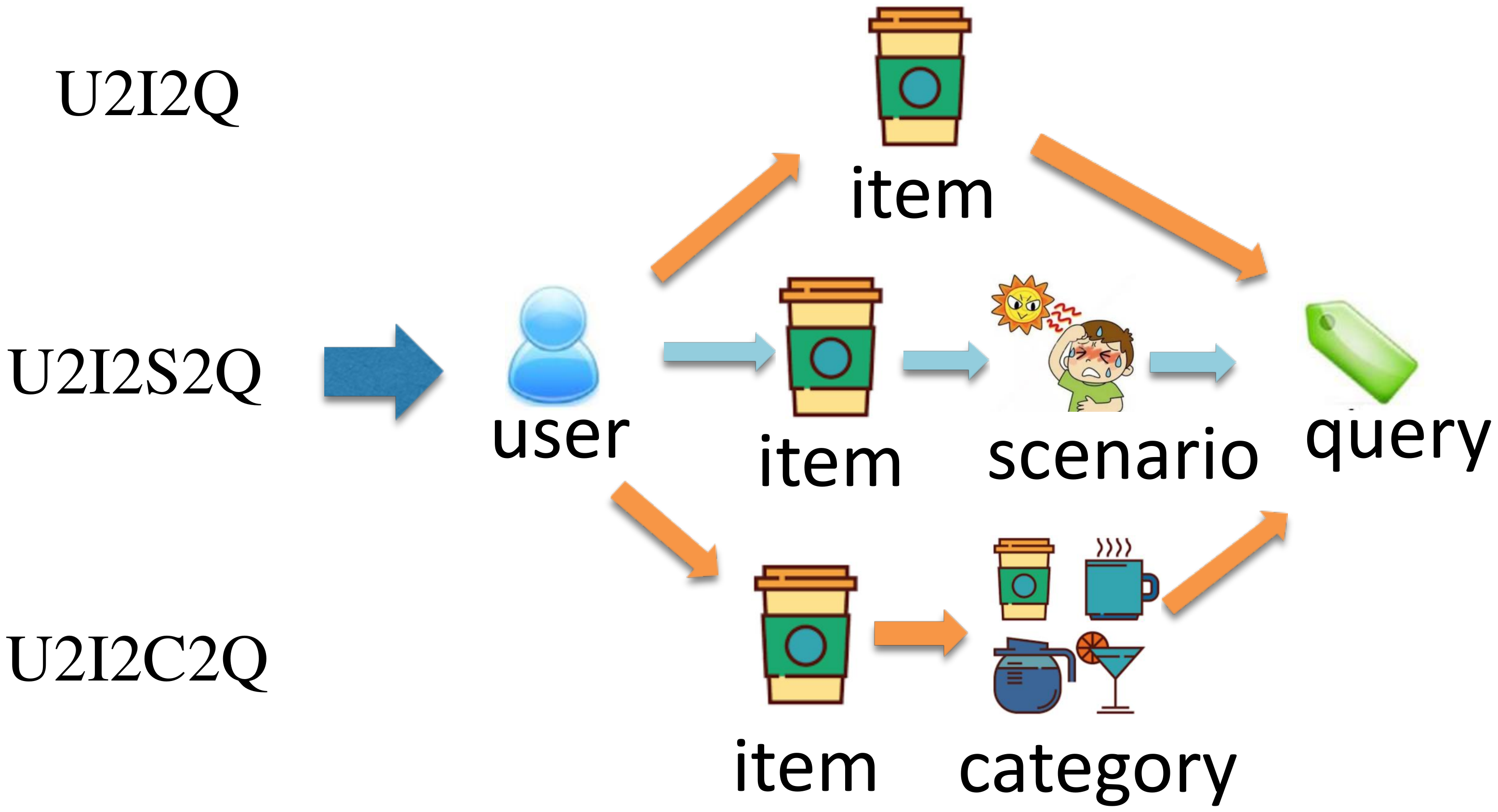}
\end{center}
   \caption{Three types of meta paths to generate query candidates.
}
\label{fig:metaPath}
\end{figure}

Candidate Generation is to efficiently generate hundreds of query candidates from the large query pool. As shown in Figure \ref{fig:metaPath}, we design three types of meta paths to generate candidates: U2I2Q, U2I2S2Q and U2I2C2Q.\vspace{10pt}

\begin{figure*}[tb]
\begin{center}
\includegraphics[scale=0.21]{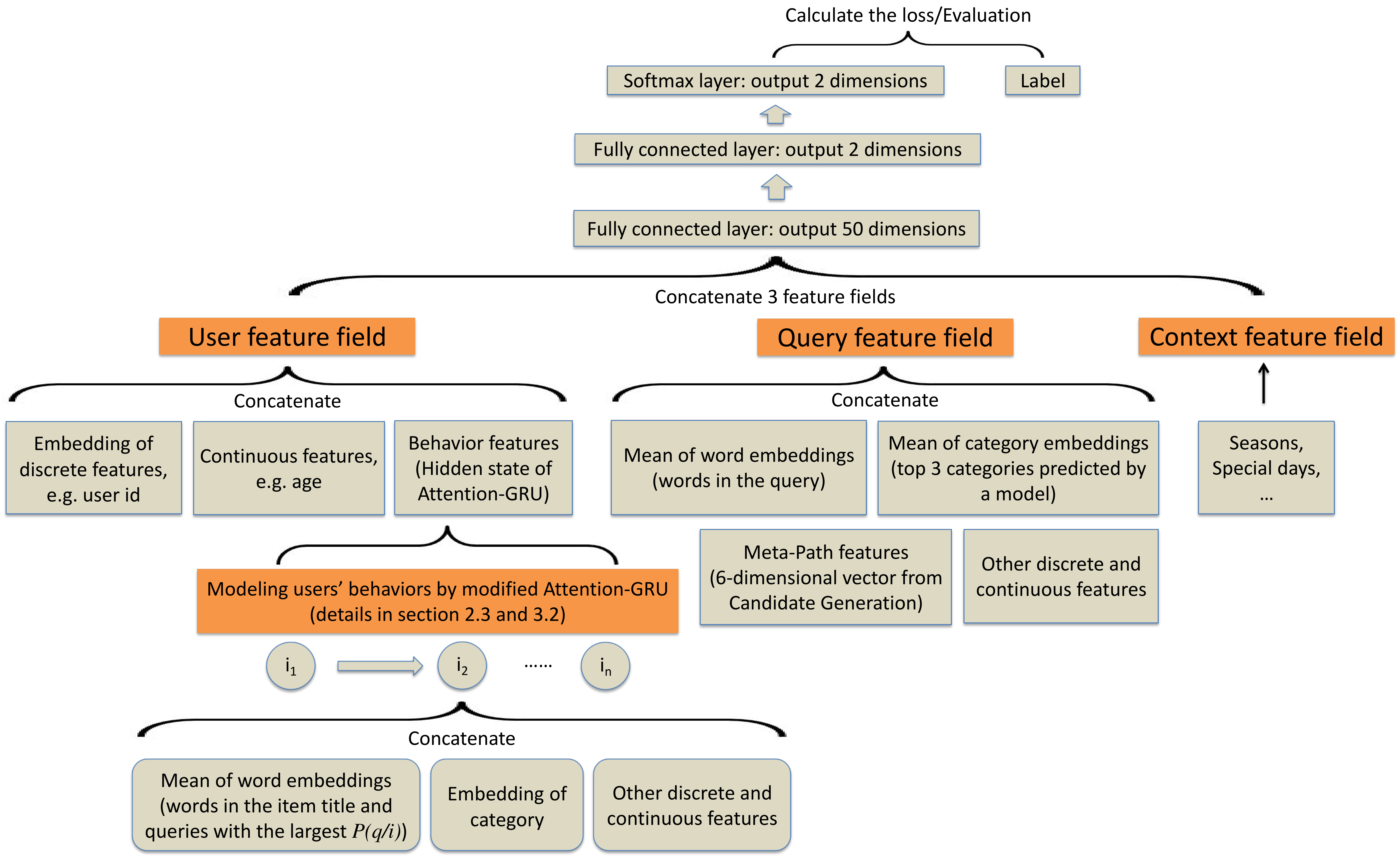}
\end{center}
   \caption{Attention-GRU based model to rank query candidates.
}
\label{fig:rankingFramework}
\end{figure*}

\noindent \textbf{\emph{U2I2Q}:} Based on search logs in previous days, we calculate the conditional probability of query $q$ given item $i$ as follows: 
\begin{equation}\label{eq:i2q}
\begin{aligned}
P(q/i) = \frac{Count(q,i)}{Count(i)},
\end{aligned}
\end{equation}
where $Count(q,i)$ is the number of records that the retrieved items contain $i$ by searching $q$, and $Count(i)$ is the number of all records that $i$ is retrieved. The insight is that, if $q$ and $i$ always co-occur in search logs, then they are closely related. For a given user $u$, it indicates that $u$ could find his preferred queries by the meta path: $p^{U2I2Q} = u \stackrel{Consume}{\longrightarrow} i \stackrel{P(q/i)}{\longrightarrow} q$.\vspace{10pt}

\noindent \textbf{\emph{U2I2S2Q}:} 
U2I2S2Q contains two sub-procedures: I2S and S2Q. I2S indicates that when user $u$ consumes item $i$, it would activate a scenario $s$, with $P(s/i)$ calculated considering $i$'s title and category. For S2Q, we derive the conditional probability of other items $i^\prime$ given $s$, i.e. $P(i^\prime/s)$, based on $i^\prime$'s titles and categories. We then obtain $P(q/s)$ with the help of $P(q/i^\prime)$ in Eq. (\ref{eq:i2q}). Finally, $u$ could find his preferred queries by the meta path: $p^{U2I2S2Q} = u \stackrel{Consume}{\longrightarrow} i \stackrel{P(s/i)}{\longrightarrow} s \stackrel{P(q/s)}{\longrightarrow} q$.\vspace{10pt}

\noindent \textbf{\emph{U2I2C2Q}:} In order to generate more query candidates, we further construct the meta path: $p^{U2I2C2Q} = u \stackrel{Consume}{\longrightarrow} i \stackrel{P(c/i)}{\longrightarrow} c \stackrel{P(q/c)}{\longrightarrow} q$, where $P(c/i) = 1$ if item $i$ belongs to category $c$. $P(q/c)$ is calculated based on the knowledge graph. 

We omit the description on how to calculate $P(s/i)$, $P(q/s)$ and $P(q/c)$ in detail since it is not the focus of this paper. $MetaScore(q)$ denotes the relation weight between $u$ and $q$ in terms of a certain type of meta path, defined as:
\begin{equation}\label{eq:MetaScore}
\small
\begin{aligned}
MetaScore(q)=\mbox{\hspace*{5cm}}\\
   \begin{cases}
   \sum\limits_{p \in U2I2Q}P(q/i) &\mbox{if $q$ is generated by U2I2Q}\\
   \sum\limits_{p \in U2I2S2Q}P(s/i) \times P(q/s) &\mbox{if $q$ is generated by U2I2S2Q}\\
   \sum\limits_{p \in U2I2C2Q}P(c/i) \times P(q/c) &\mbox{if $q$ is generated by U2I2C2Q}
   \end{cases}.
\end{aligned}
\end{equation}
We collect $u$'s recently consumed (\emph{clicked}, \emph{purchased}, \emph{favored}, \emph{added-to-cart}) items as $i$. The weight of a meta path is the product of different conditional probabilities along the meta path. The probability product is used because, we assume that in Meta-Path, the observation on one node only depends on its previous node, i.e. following the first-order Markov assumption. Taking a certain meta path in U2I2S2Q as an example, its probability of occurrence $P(i,s,q)$ (item $i$, scenario $s$ and query $q$ are along the meta path) is equal to $P(i) \times P(s/i) \times P(q/s)$ according to the Markov assumption. $P(i)$ is assumed to be the same for all items recently consumed by the target user. Therefore, $P(s/i) \times P(q/s)$ in Eq. (\ref{eq:MetaScore}) is proportional to $P(i,s,q)$. The other types of meta paths can be similarly analyzed. There may exist multiple meta paths that link $u$ and $q$, thus we sum their weights to obtain $MetaScore(q)$.

For a query candidate, we represent the output of Meta-Path by a 6-dimensional vector: $[Type_1, Score_1, Type_2, Score_2, Type_3, Score_3]$. If there exists at least one meta path belonging to U2I2Q that links $u$ and $q$, then we have $Type_1$ = 1 and $Score_1$ is $MetaScore(q)$ with respect to U2I2Q. Otherwise, we have $Type_1 = 0$ and $Score_1 = 0$. $Type_2$, $Score_2$ are for U2I2S2Q and $Type_3$, $Score_3$ are for U2I2C2Q.

We calculate $i \rightarrow \dots \rightarrow q$ relations, save $i$'s top-$k$ queries with the largest conditional probabilities and build index on $i$ for each type of meta path offline. Then in online recommendation, given user $u$, query candidates can be efficiently generated according to $u$'s consumed items by indexing. In our system, we set the maximum number of query candidates for each type of meta path to be 200. In this way, the total number of query candidates for each user request is controlled to be less than or equal to 600 (since we have three types of meta paths). Note that other relations apart from the above three can also be incorporated in our Meta-Path model. Compared to the embedding model for Candidate Generation in \cite{covington:2016deep}, ours exploits heterogeneous relations to generate candidates, is more explainable and is easier to implement.

\subsection{Attention-GRU Based Model for Ranking}
The Ranking stage is to effectively rank query candidates and top queries are then recommended. We formulate it as a point-wise ranking task. Specifically, a classifier is learned by exploiting various information in e-commerce websites. Then given user $u$, his probability of preferring each query is predicted by the classifier and queries are ranked by their probabilities. As shown in Figure \ref{fig:rankingFramework}, there are $3$
feature fields in our ranking model: user features, query features and context features. \vspace{10pt}

\noindent \textbf{\emph{user features}:} This feature field contains users' discrete features (e.g. user id), continuous features (e.g. age) and behavior features. We encode discrete features by embedding. Taking user id as an example, the user set is denoted by $U = \{u_1, u_2, \cdots, u_N\}$. $u_k$'s one-hot vector is defined as $o_k\in \{0,1\}^{N\times 1}$, with the $k$-th entry equal to  $1$ and the other entries equal to $0$. Then we obtain $u_k$'s embedding $e_k$ as:
\begin{equation}\label{eq:embedding for discrete features}
\begin{aligned}
&e_k = \mathbf{E}\times o_k.
\end{aligned}
\end{equation}
$\mathbf{E} \in R^{D\times N}$ contains the embeddings of all users, which is learned from training. $o_k$ is used to look up the embedding of $u_k$ from $\mathbf{E}$. Behavior features are represented by the hidden state of our modified Attention-GRU, considering the text information, categories, other discrete and continuous features of users' consumed items. The representation of text is not the focus of this paper, thus we simply represent it by the mean of word embeddings. Categories are discrete features, thus they are directly encoded by embedding. \vspace{10pt}

\noindent \textbf{\emph{query features}:} This feature field contains the text information, categories, Meta-Path features, other discrete and continuous features of the query. Note that there is no explicit category information on queries, thus we learn a model to predict the top 3 categories for each query, and encode them by the mean of category embeddings. Meta-Path features are the 6-dimensional vector obtained from Candidate Generation. \vspace{10pt}

\noindent \textbf{\emph{context features}:} This feature field contains seasons, special days, etc., which also influence the ranking performance.

The concatenation of these feature fields are the input of a 3-layer neural network (2 fully connected layers and 1 softmax layer). Loss/evaluation is calculated based on its output and the label. Some features capture the system bias, e.g. for the feature of special days, users are more likely to click queries in Shopping Festivals than in normal days. Some features capture the bias on users or queries, e.g. for the feature of user id, some active users prefer to click most queries. Similarly, queries with popular categories tend to be clicked by most users. Other features capture users' personalized preferences on queries, e.g. the text information of a user's consumed items models his preference by text. Then if the query also contains similar text, he would probably prefer this query. Different from linear models, neural network used in our framework not only captures the bias of unary features (i.e. the bias on system, users and items), but also well models the interaction among features (e.g. users' personalized preferences on queries) by non-linear activation functions.

Our ranking model is inspired by the Wide\&Deep model \cite{cheng:2016wide}. 
Obtaining behaviour features from Attention-GRU in Figure \ref{fig:rankingFramework} can be regarded as the deep component in \cite{cheng:2016wide} while the other features construct the wide component. However, our model and the Wide\&Deep model in \cite{cheng:2016wide} share some key differences. For the deep component, we use the hidden state of our modified Attention-GRU as features while \cite{cheng:2016wide} regards the final score of Deep Neural Network (DNN) \cite{hinton:2012deep} as the feature. Thus our model exploits users' sequential behaviors for recommendation while \cite{cheng:2016wide} cannot. For the combination of the wide and deep components, \cite{cheng:2016wide} combines them by logistic regression while we use a 3-layer neural network to model them, so that the interaction among features in wide and deep components can be better captured.

\begin{figure}[tb]
\begin{center}
\includegraphics[scale=0.18]{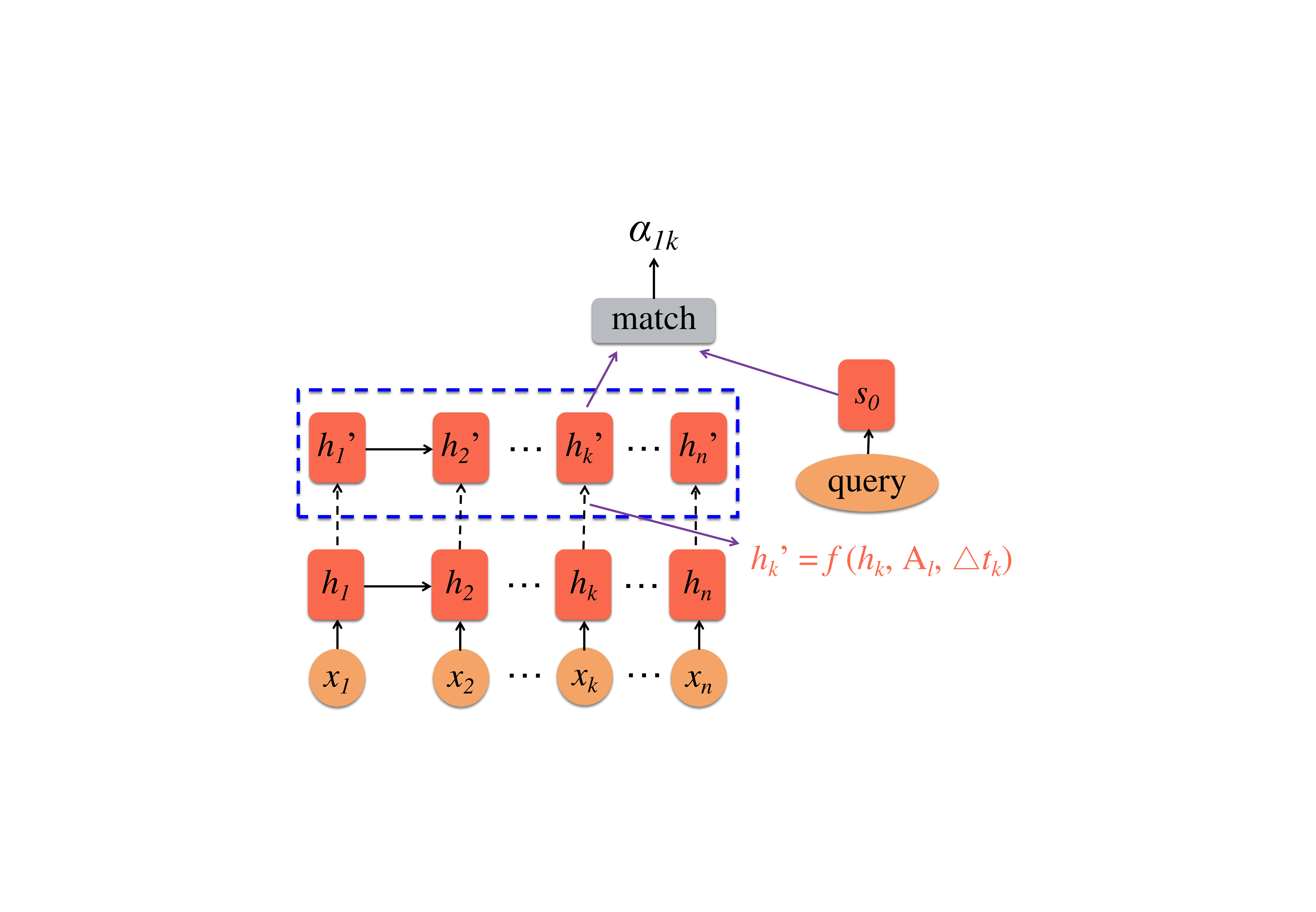}
\end{center}
   \caption{Modified attention schema considering action types and time decay.
}
\label{fig:calculation of attention weight}
\end{figure}

\vspace{10pt}
\noindent \textbf{\emph{Modified Attention-GRU}}\vspace{3pt}\\
Now we describe our modified Attention-GRU in detail. Following the denotations in section 2.3, $x_k$ corresponds to item $i_k$ in Figure \ref{fig:rankingFramework}. $T$ is equal to $1$, thus we have $m = 1$ in Eq. $(1)\sim(4)$. $y_0$ corresponds to the query. $s_0$ is the hidden state of $y_0$. We use bidirectional GRU to output $h=(h_1,\cdots, h_n)$, with $h_k$ defined as:
\begin{align}
&\overrightarrow{h_k} = GRU(x_k, \overrightarrow{h_{k-1}})\label{eq:bidirectional GRU1},\\
&\overleftarrow{h_k} = GRU(x_k, \overleftarrow{h_{k+1}})\label{eq:bidirectional GRU2},\\
&h_k = [\overrightarrow{h_k}, \overleftarrow{h_k}]\label{eq:bidirectional GRU3}.
\end{align}
The concatenation of $h_k$ (containing the information of $i_k$) and $s_0$ (containing the query information) is the input of a neural network, and its output is the attention weight $\alpha_{1k}$. Finally, we use $s_1$ in Eq. (\ref{eq:Attention-GRU3}) to represent behavior features in Figure \ref{fig:rankingFramework}. Our modifications on Attention-GRU are based on the following two motivations:
\begin{itemize}
\item Different action types (\emph{click}, \emph{purchase}, \emph{favor}, \emph{add-to-cart}) reflect users' different preferences on items, e.g. generally a user purchasing an item indicates he is more interested in the item than if he clicks it. The design of attention weight should consider different action types.
\item The earlier an action happens, the less it affects query recommendation. Thus the time decay of different actions should also be modeled in attention weight.
\end{itemize}
Following these motivations, as shown in Figure \ref{fig:calculation of attention weight}, we replace $h$ in Eq. (\ref{eq:Attention-GRU1}) with $h^\prime=(h_1^\prime,\cdots, h_n^\prime)$, where $h_k^\prime$ is defined as:
\begin{align}
&h_k^{tmp} = \mathbf{A}_{l}\times h_k,\label{eq:modified GRU1}\\
&h_k^\prime = h_k^{tmp} \otimes \triangle t_k^{\epsilon}, \mbox{~~~$s.t.$~~~} \epsilon \leq 0. \label{eq:modified GRU2}
\end{align}
Suppose we have $h_k\in R^{d\times 1}$, then we define $\mathbf{A}_{l}\in R^{d\times d}$, where $l$ indicates whether $i_k$ is \emph{clicked} ($l = 1$), \emph{purchased} ($l = 2$) , \emph{favored} ($l = 3$) or \emph{added-to-cart} ($l = 4$). We use matrix multiplication to model different action types in Eq. (\ref{eq:modified GRU1}), since in this way the interaction between the action type and $i_k$ is explicitly captured. $\triangle t_k$ is the time interval between the time when the action on $i_k$ happens and the time for query recommendation. $\epsilon$ is an exponential decay on $\triangle t_k$. $\otimes$ indicates that each entry in $h_k^{tmp}$ is multiplied by $\triangle t_k^{\epsilon}$. Constraint $\epsilon \leq 0$ in Eq. (\ref{eq:modified GRU2}) ensures that the earlier an action on $i_k$ happens, the smaller $h_k^\prime$ will be. Correspondingly, $i_k$'s influence on query recommendation will also be smaller. $\mathbf{A}_{l}$ and $\epsilon$ are learned from training. Constraint $\epsilon \leq 0$ is handled by using the projection operator \cite{rakhlin:2012making}, i.e. if we have $\epsilon > 0$ during training iterations, we reset $\epsilon = 0$. Note that our two modifications can be generalized to other attention models, by replacing their $h_k$ with our $h_k^\prime$ as shown in Eq. (\ref{eq:modified GRU1}) and Eq. (\ref{eq:modified GRU2}).

\section{Offline Experiments}
\subsection{Dataset}
We firstly deploy our interactive RS in Taobao, using a heuristic ranking method (we cannot use our ranking model since no training data is available yet) to recommend queries. Then one day of data is collected and preprocessed for training and testing. For each instance $<$\emph{uid, qid, label}$>$, \emph{label} $ = 1$ indicates that query \emph{qid} is clicked by user \emph{uid} and \emph{label} $ = 0$ means that \emph{qid} is shown to but not clicked by \emph{uid}. In addition, user features (behavior features are constructed by \emph{uid}'s recent $100$ actions), query features and context features are correspondingly collected. Finally, we have $3,201,231$ users, $1,464,410$ queries and $12,897,055$ instances. $80\%$ instances are randomly selected for training and the remaining $20\%$ are for testing.

\subsection{Compared Models and Evaluations}
Our ranking model is compared with the following baselines.\vspace{10pt}\\
\noindent \textbf{\emph{Q\&R}:} We denote the question ranking model in state-of-the-art interactive RS \cite{christakopoulou:2018q} (replace its topics with queries) as Q\&R.\\
\noindent \textbf{\emph{FTRL}:} FTRL \cite{mcmahan:2013ad} is a linear model, with no feature interaction. Behavior features in Figure \ref{fig:rankingFramework} are not used since they cannot be easily incorporated in FTRL. We manually design some interactive features and behavior features for FTRL to ensure more fair comparisons.\\
\noindent \textbf{\emph{Wide\&Deep}:} Wide\&Deep learning \cite{cheng:2016wide} is a popular learning framework in industry. Here we compare our model with its famous version described in \cite{cheng:2016wide}. Based on the discrete and continuous features in Figure \ref{fig:rankingFramework}, we construct raw input features and transformed
features for the wide component, and use DNN to generate real-valued vectors for the deep component. Finally, the wide and deep components are combined by a logistic regression model. Refer to \cite{cheng:2016wide} for more details.

The remaining baselines replace our modified Attention-GRU
with other RNN structures. We use the corresponding RNN structures to denote these baselines for simplicity.\\
\noindent \textbf{\emph{GRU}:} GRU \cite{cho:2014learning} is one of the best RNN architectures. Thus it is selected to represent the original RNN structures, with timestamps and action types not considered.\\
\noindent \textbf{\emph{Attention-GRU}:} Similarly, we choose Attention-GRU \cite{chorowski:2015attention} to represent RNN structures with the attention schema. Timestamps and action types are not considered, either.\\
\noindent \textbf{\emph{Time-LSTM}:} Time-LSTM \cite{zhu:2017next} has achieved state-of-the-art performance for sequential behavior modeling when timestamps exist while action types are not known. We use its publicly available python implementation\footnote{https://github.com/DarryO/time\_lstm}.\\
\noindent \textbf{\emph{Attention-GRU-3M}:} Attention-GRU-3M \cite{zhu:2018brand} considers timestamps and action types in sequential behavior modeling. However, its time intervals are calculated between neighbor actions, which are different from ours. In addition, we propose two important modifications on the attention schema to better model timestamps and action types. We use its publicly available python implementation\footnote{https://github.com/zyody/Attention-GRU-3M}.

The number of units is set to 256 for all RNN-based structures. The other hyper-parameters in all models are tuned via cross-validation or set as in the original paper. We evaluate the performance of different models by AUC and F1 score \cite{kim:2013nonparametric}.

\begin{table}[t]
\centering
\begin{tabular}{|c|c|c|c|c|c|}
\hline
 \multicolumn{2}{|c|}{}&\multicolumn{2}{|c|}{50\% training data}&\multicolumn{2}{|c|}{100\% training data}\\ \hline
 \multicolumn{2}{|c|}{}&AUC&F1&AUC&F1\\ \hline
\multicolumn{2}{|c|}{Q\&R}&0.651&0.638&0.671&0.648\\ \hline
\multicolumn{2}{|c|}{FTRL}&0.647&0.635&0.669&0.644\\ \hline
\multicolumn{2}{|c|}{Wide\&Deep}&0.651&0.639&0.673&0.650\\ \hline
\multicolumn{2}{|c|}{GRU}&0.655&0.642&0.676&0.653\\ \hline
\multicolumn{2}{|c|}{Attention-GRU}&0.661&0.649&0.683&0.662\\ \hline
\multicolumn{2}{|c|}{Time-LSTM}&0.664&0.651&0.684&0.666\\ \hline
\multicolumn{2}{|c|}{Attention-GRU-3M}&0.671&0.657&0.690&0.674\\ \hline
\multicolumn{2}{|c|}{Our Model}&\textbf{0.682}$^*$&\textbf{0.667}$^*$&\textbf{0.699}$^*$&\textbf{0.685}$^*$\\ \hline
\end{tabular}
\vspace*{15pt}
\caption{Model Comparison (* indicates statistical significance at $p < 0.01$ compared to the second best.)}
\label{table:Method Comparison}
\end{table}
\subsection{Results and Discussions}
\subsubsection{Model Comparison}
As shown in Table \ref{table:Method Comparison}, our proposed model significantly outperforms all baselines. In comparison, Q\&R predicts the query conditioned on a sequence of consumed items, with features not carefully engineered. Moreover, it uses GRU, which is inferior compared to our modified Attention-GRU. FTRL is a linear model, with interactive features and behavior features manually designed. However, it is difficult for FTRL to capture complex feature interactions and well model sequential behaviors. The DNN used in Wide\&Deep fails to model sequential behaviors, either. In addition, its combination model, i.e. logistic regression, is linear, which cannot well capture the interaction among features in the wide and deep components. Attention-GRU performs better than GRU due to the attention schema, but it is worse than our model, which demonstrates that adding timestamps and action types in behavior modeling can improve the ranking performance. Time-LSTM fails to distinguish different action types. Attention-GRU-3M considers timestamps and action types, but still performs worse than our model, which proves the advantage on how we model timestamps and action types.

\begin{table}[t]
\centering
\begin{tabular}{|c|c|c|c|c|c|c|c|c|}
\hline
&AUC&Gain\\ \hline
Random prediction&0.500&-\\ \hline
Add categories of the query&0.568&+0.068\\ \hline
Add categories of sequential items&0.604&+0.036\\ \hline
Add texts of sequential items and the query&0.620&+0.016\\ \hline
Add all features in Figure \ref{fig:rankingFramework}&0.683&+0.063\\ \hline
Add the modification in Eq. (\ref{eq:modified GRU1})&0.689&+0.006\\ \hline
Add the modification in Eq. (\ref{eq:modified GRU2})&0.699&+0.010\\ \hline
\end{tabular}
\vspace*{15pt}
\caption{Evaluation on Different Ingredients of Our Ranking Model (\emph{Gain} represents the improvement of current AUC compared to the previous one.)}
\label{table:Evaluation on Different Ingredients in Our Framework}
\end{table}
\subsubsection{Contributions of Different Ingredients}
We further conduct experiments to evaluate the contributions of different ingredients in our ranking model. As shown in Table \ref{table:Evaluation on Different Ingredients in Our Framework}, random prediction achieves an AUC of 0.500. If we only use the query's categories as input, AUC reaches 0.568, which indicates the popularity of query's categories is useful for query recommendation. When the categories of sequential items are added, AUC increases to 0.604, demonstrating that item categories and their sequential information contribute to performance improvement. Adding the text information of sequential items and the query further increases AUC to 0.620, which verifies the effectiveness of texts. Attention-GRU with all features in Figure \ref{fig:rankingFramework} achieves an AUC of 0.683, proving the usefulness of other features. Finally, our modifications on attention schema, i.e. Eq. (\ref{eq:modified GRU1}) and Eq. (\ref{eq:modified GRU2}), obtain the AUC gain of 0.006 and 0.010, respectively, which verifies the effectiveness of our two modifications.

\begin{figure}[t!]
\begin{center}
\subfigure[The original recommendation setting.]{\includegraphics[scale=0.2]{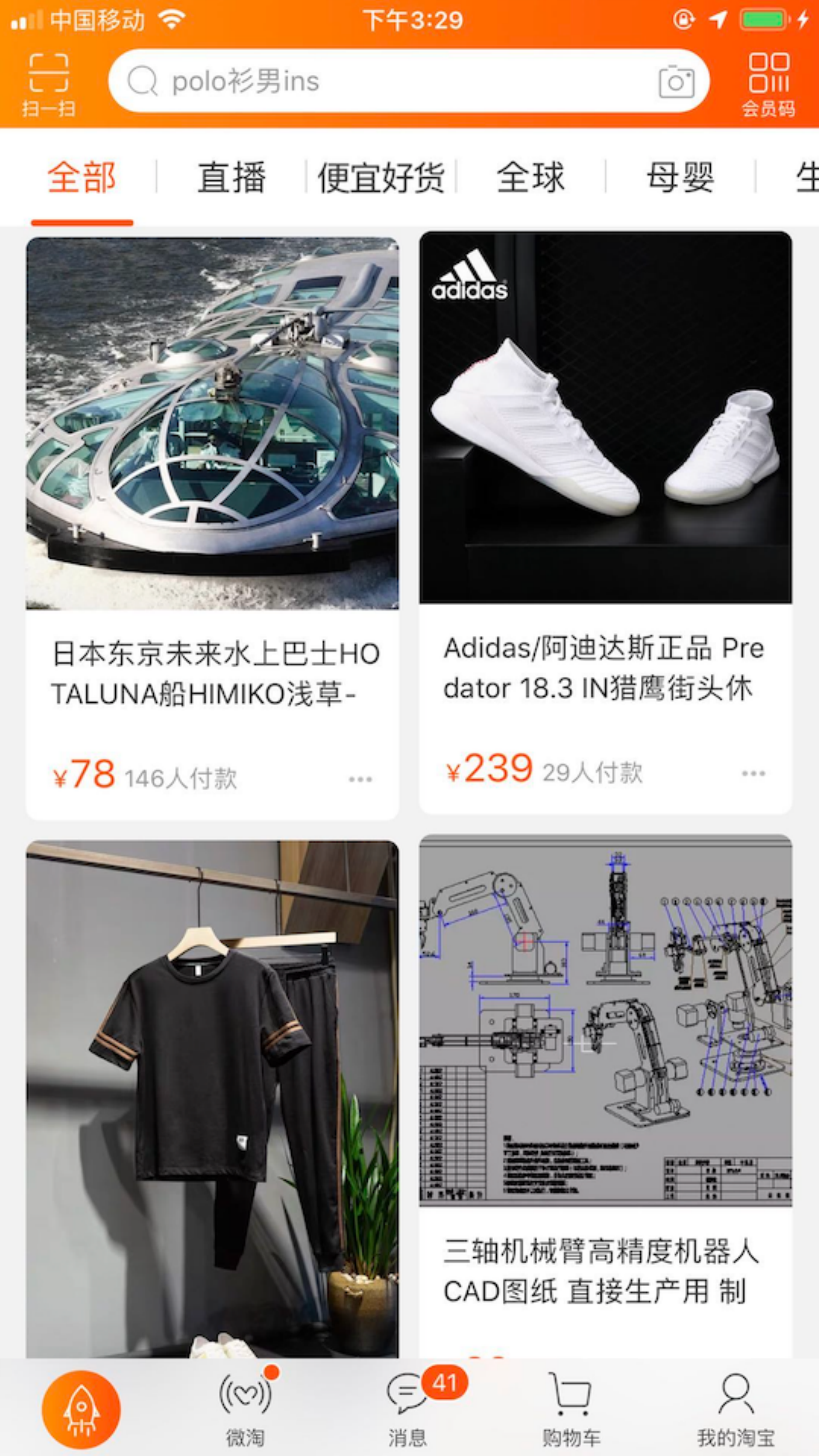}}
\hspace{12pt}
\subfigure[Occasionally presenting the user interface of our interactive RS, highlighted by the red rectangle.]{\includegraphics[scale=0.2]{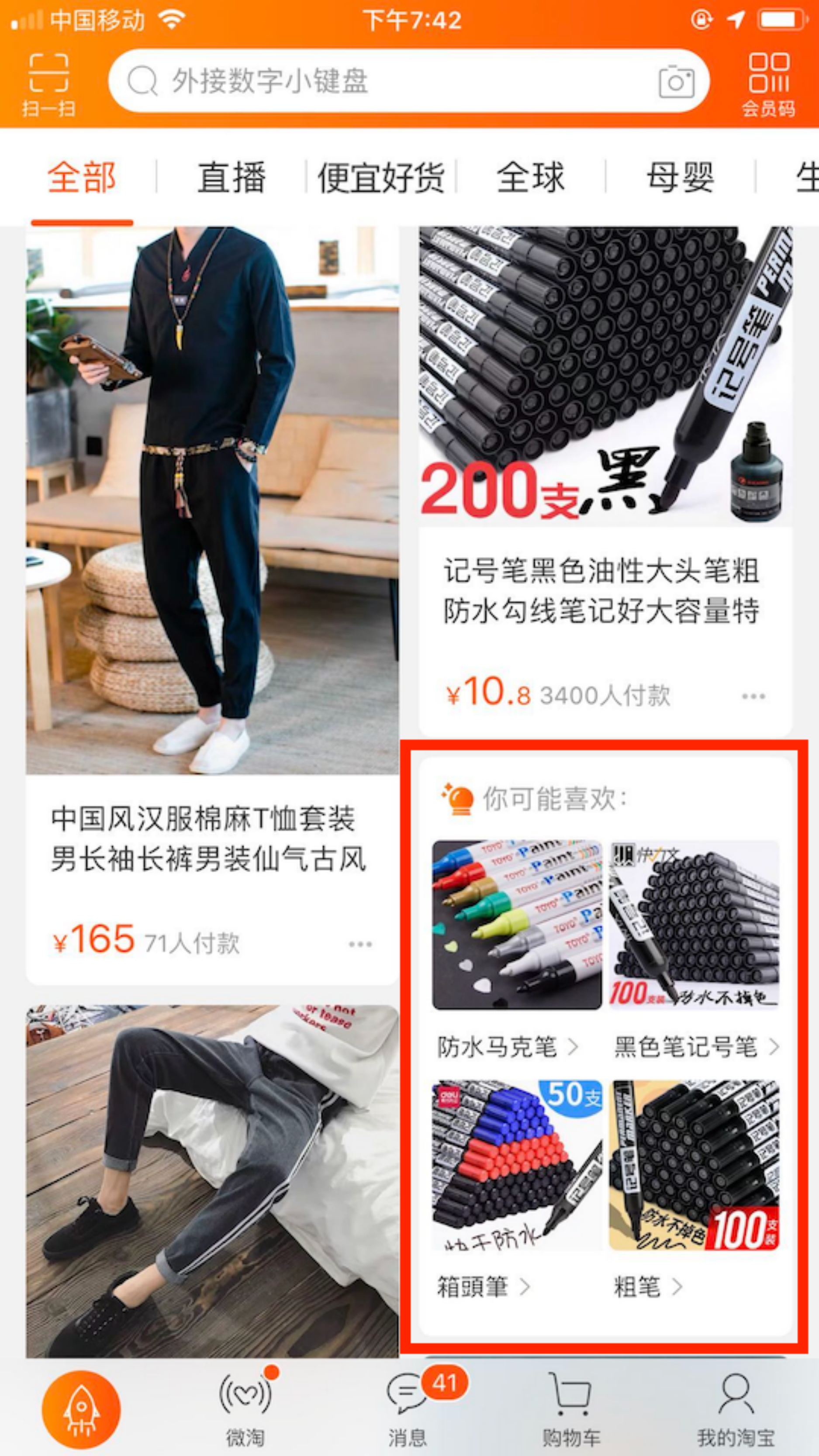}}
\end{center}
   \caption{A/B test setting. (a) is the original setting. Our interactive RS is added in (b).
}
\label{fig:AB test}
\end{figure}

\section{Online Experiments}
We now test real users' response to our interactive RS in Taobao. In this large-scale platform, there are over $10^{10}$ \emph{impressions} and about $6\times 10^8$ \emph{clicks} on over $10^7$ items from nearly $7\times 10^7$ customers within one normal day. A standard A/B test is conducted online, where one adopts the original recommendation setting as shown in Figure \ref{fig:AB test} (a) and the other occasionally presents the user interface of our interactive RS as shown in Figure \ref{fig:AB test} (b). The appearance of interactive user interfaces is controlled by an intention model, and often happens after the user clicks some items and returns back to the homepage of Taobao App. See https://v.qq.com/x/page/s0833tkp1uo.html on how it works online. The same number (about $7\times 10^6$ per day) of users are randomly selected for each setting. We perform the online experiments for five days, and the average impression number on items (denoted as \emph{Impression}), click number on items (denoted as \emph{Click}) and Gross Merchandise Volume (denoted as \emph{GMV}) per day are reported.

As shown in Table \ref{table:Performance of Online Experiment}, by adding our interactive RS, \emph{Impression}, \emph{Click} and \emph{GMV} are all improved. A higher \emph{Impression} and \emph{Click} indicates that users are more willing to browse and click items in our interactive RS. The improvement of \emph{GMV} is larger, because based on users' feedback, the system can well learn users' shopping needs and then satisfy them, which would lead to much more purchases. Considering the platform's traffic, 2.93\% improvement on \emph{GMV} would result in a significant boost in revenue. 

In our platform, if we increase the number of query candidates generated by the Candidate Generation stage, both of users' Click-Through Rate (denoted as \emph{CTR}) on queries and the \emph{response time} will increase. When the number is larger than $900$, it will exceed the system's constraint on \emph{response time}. \emph{CTR} increases because the model in the Ranking stage is more effective than the one in the Candidate Generation stage. Some users' preferred queries may have low scores in Candidate Generation but will be correctly ranked in Ranking. Therefore, users are more likely to find their preferred queries with a larger candidate set and thus result in a higher \emph{CTR}. \emph{response time} increases because more query candidates lead to more predictions in the Ranking stage, which is obviously more time consuming. Hence, we should carefully set the number of query candidates to trade off the efficiency and effectiveness. This also verifies the necessary of the Candidate Generation stage, since it is impossible to rank millions of queries in the Ranking stage within an acceptable time. 

Our interactive RS has already gone into production on Taobao since Nov. 11, 2018, with about $4.5\times 10^7$ active users per day. Users will see the interactive user interfaces after clicking some items and returning back to the homepage of Taobao App. The ranking model is daily updated, which is initialized by the parameters in the previous day and fine-tuned with the data obtained in the new day. In this way, the model can not only remember old data but also continuously fit the latest data to achieve better results.

\begin{table}[tb]
\centering
\begin{tabular}{|c|c|c|c|}
\hline
 &Impression&Click&GMV (CNY)\\ \hline
 Original setting&1448730074&62956757&9261422\\ \hline
 Add our interactive RS&1469132334&63361783&9532320\\ \hline
 Improvement&1.41\%&0.64\%&2.93\%\\ \hline
\end{tabular}
\vspace*{15pt}
\caption{Results of Online Experiments (\emph{Improvement} is a relative growth of \emph{Add our interactive RS} compared to \emph{Original setting}, e.g. $2.93\%\approx (9532320-9261422)/9261422$.)}
\label{table:Performance of Online Experiment}
\end{table}

\section{Conclusion}
In this paper, we propose a query-based interactive RS, which can accurately generate personalized questions and recommend items closely-related to users' feedback. To ensure high efficiency and remarkable effectiveness, we propose a model based on Meta-Path for the Candidate Generation stage and an adapted Attention-GRU model for the Ranking stage. Offline and online experiments verify the effectiveness of our interactive RS. In future work, we will explore more about the \emph{Item Recommendation} module. Furthermore, we would try to generate questions by other objects (e.g. videos), besides queries, and recommend items according to users' behaviors on these objects to improve the user experience in interactive RS.

%

%
\bibliographystyle{ACM-Reference-Format}
\bibliography{cikm19}

%
\end{document}